\documentclass[journal ]{new-aiaa}
\usepackage[utf8]{inputenc}
\usepackage{textcomp}

\usepackage{graphicx}
\usepackage{amsmath}
\usepackage{mathtools}
\usepackage[version=4]{mhchem}
\usepackage{siunitx}
\usepackage{longtable,tabularx}
\usepackage{algorithm}
\usepackage[noend]{algpseudocode}

\newcommand{\hide}[1]{}

\usepackage{xcolor}
\definecolor{cardinal}{rgb}{0.0, 0.8, 0.6}

\newcommand{\bx}{\textbf{\emph{x}}}

\newcommand{\ba}{\textbf{\emph{a}}}

\newcommand{\bl}{\boldsymbol{\ell}}

\newcommand{\bg}{\textbf{\emph{g}}}
\newcommand{\bt}{\textbf{\emph{t}}}
\newcommand{\bn}{\textbf{\emph{n}}}

\newcommand{\bm}{\textbf{\emph{m}}}
\newcommand{\br}{\textbf{\emph{r}}}

\newcommand{\bu}{\textbf{\emph{u}}}

\newcommand{\by}{\textbf{\emph{y}}}

\newcommand{\bp}{\textbf{\emph{p}}}

\newcommand{\bA}{\textbf{\emph{A}}}
\newcommand{\bC}{\textbf{\emph{C}}}

\newcommand{\bI}{\textbf{\emph{I}}}
\newcommand{\bP}{\textbf{\emph{P}}}
\newcommand{\bK}{\textbf{\emph{K}}}
\newcommand{\bL}{\textbf{\emph{L}}}
\newcommand{\bH}{\textbf{\emph{H}}}
\newcommand{\bR}{\textbf{\emph{R}}}

\newcommand{\bbN}{\textbf{\emph{N}}}
\newcommand{\bk}{\textbf{\emph{k}}}
\newcommand{\bB}{\textbf{\emph{B}}}
\newcommand{\bT}{\textbf{\emph{T}}}

\newcommand{\bM}{\textbf{\emph{M}}}
\newcommand{\bS}{\textbf{\emph{S}}}

\newcommand{\bxi}{\boldsymbol{\xi}}

\newcommand{\brho}{\boldsymbol{\rho}}

\newcommand{\blam}{\boldsymbol{\lambda}}

\newcommand{\PP}{{\mathbb P}}
\newcommand{\RR}{{\mathbb R}}

\title{Pole Estimation and Optical Navigation using Circle of Latitude Projections}

\author{John A. Christian \footnote{Associate Professor, Guggenheim School of Aerospace Engineering. Associate Fellow AIAA.}}
\affil{Georgia Institute of Technology, Atlanta, GA 30332}

\begin{document}

\maketitle

\begin{abstract}
Images of both rotating celestial bodies (e.g., asteroids) and spheroidal planets with banded atmospheres (e.g., Jupiter) can contain features that are well-modeled as a circle of latitude (CoL). The projections of these CoLs appear as ellipses in images collected by cameras or telescopes onboard exploration spacecraft. This work shows how CoL projections may be used to determine the pole orientation and covariance for a spinning asteroid. In the case of a known planet modeled as an oblate spheroid, it is shown how similar CoL projections may be used for spacecraft localization. These methods are developed using the principles of projective geometry. Numerical results are provided for simulated images of asteroid Bennu (for pole orientation) and of Jupiter (for spacecraft localization).
\end{abstract}

\section{Introduction}

\begin{figure}[b!]
    	\centering
    	\includegraphics[width=0.25\textwidth]{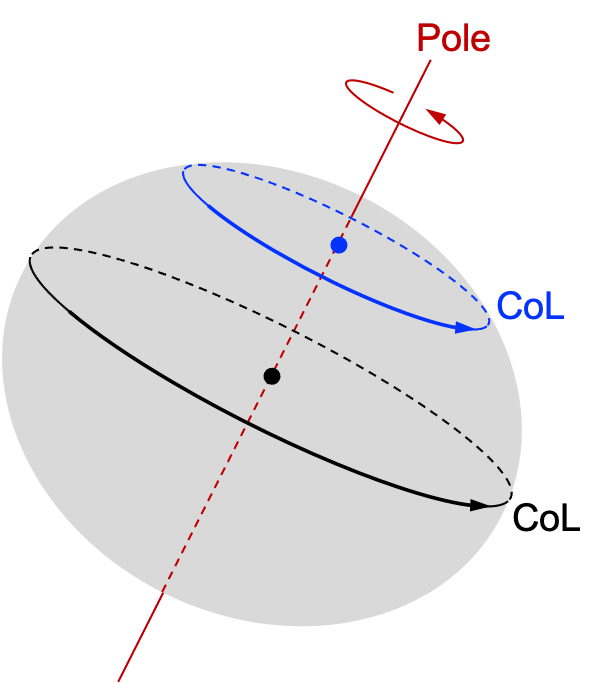}
    	\caption{A circle of latitude (CoL) on a rotating celestial body lies in a plane perpendicular to the pole.}
    	\label{fig:PoleGeom}
\end{figure}

It is common to encounter celestial bodies rotating about their axis of largest moment of inertia. 
This naturally stable configuration describes the planets in our Solar System as well as many small bodies (e.g., asteroids, minor planets). Indeed, principal axis rotation is the norm for asteroids, with significant departures from this state primarily occurring for slowly rotating bodies \cite{Pravec:2000}.
Therefore, given the prevalence of principal axis rotators in the Solar System, it is interesting to consider how these dynamics might be exploited for shape modeling, navigation, and other practical purposes. Although principal axis rotation leads to a variety of helpful geometric properties, this work focuses on just one particular insight: the inertial path of any single point on the rotating body's surface must trace out a circle relative to the body's center. This is true regardless of the body's actual three-dimensioanl (3-D) shape. Such circles are commonly referred to as a \emph{circle of latitude} (CoL). The body's axis of rotation (i.e., the \emph{pole}) is perpendicular to the plane of each CoL and passes through each circle's center (see Fig.~\ref{fig:PoleGeom}). These CoLs project to ellipses in images collected by cameras (or telescopes), and the shape and orientation of these image ellipses may be used to infer the orientation of the observed body's pole. Moreover, the relative arrangement and sizes of many elliptical CoL projections may be used to constrain the body's shape to an unknown global scale.

This work was inspired by a number of recent studies seeking to estimate the pole of an unknown spinning asteroid using the elliptical projection of CoLs \cite{Kuppa:2022a,Kuppa:2022b}. The related problem of estimating the latitude of a spacecraft from an image of constant-latitude bands in Jupiter's atmosphere has also been of recent interest \cite{Rivera:2020,Rivera:2022,Rivera:2023}. As compared to these prior works, this manuscript aims to (1) generalize the scenarios under which the pole may be estimated from the perspective projection of CoLs and (2) considerably expand the geometric information that might be extracted from such features. 

The recent work of Kuppa, McMahon, and  Dietrich \cite{Kuppa:2022a,Kuppa:2022b} and of Danas Rivera and Peck \cite{Rivera:2020,Rivera:2022,Rivera:2023} provide excellent insights into the image processing required to localize the elliptical features in an image, and this piece of the problem is not discussed further here. These prior works, however, all make a number of limiting assumptions in the development of their navigation or reconstruction algorithms. In particular, they assume that the camera is pointed towards the center of the body and that the distance to the body is known. Neither of these assumptions are necessary to determine pole orientation or recover shape to scale. Although knowledge of the body's distance would remove the scale ambiguity, it is otherwise not needed. In addition to the assumptions on pointing and distance, the algorithms in these earlier works also require other pieces of information (or make other assumptions) that are not necessary---though the specifics vary from algorithm to algorithm. As is shown herein, all that is needed to determine the pole are the elliptical projections of CoLs in images from a calibrated camera. It does not matter where the observed CoLs lie on the body's surface nor does it matter where the camera is pointed or located relative to the body. Moreover, there is no need to correspond the observed CoLs with any known point, feature, or latitude band on the observed body. 

In addition to enhanced algorithmic generality, this work introduces a number of new capabilities, including an analytic expression for estimated pole covariance, a procedure for shape reconstruction (to scale), and a means for resolving the unknown global scale in many practical cases. The advancements described here are made possible by transferring classic computer vision results on pose from circles \cite{Shiu:1989,Dhome:1990,Forsyth:1991,Kanatani:1993,Christian:2017AAS} to the problem of pole estimation, along with some novel application-specific improvements.

\section{Perspective Projection of Circles}
Conics project to conics under the action of perspective projection \cite{Hartley:2003}. In this case, the observed conic is a circle of latitude that (in practice) almost always projects to an ellipse in an image. The geometry of circle projections is now briefly reviewed.

Consider a CoL with radius $R$. Define a coordinate frame $L$ with its origin at the circle center and the $Z$-axis perpendicular to the plane of the circle (i.e., along the pole). Thus, the $X$-axis and $Y$-axis span the plane of the circle and any two-dimensional (2-D) point $\{X,Y\}$ within the $Z=0$ plane belonging to the circle must satisfy the implicit equation
\begin{equation}
    \label{eq:CircleSimple0}
    X^2 + Y^2 - R^2 = 0
\end{equation}
Defining the $3 \times 3$ matrix $\bC_L$ of arbitrary scale,
\begin{equation}
   \bC_L \propto \begin{bmatrix}
        1 & 0 & 0 \\ 0 & 1 & 0 \\ 0 & 0 & -R^2
   \end{bmatrix}
\end{equation}
one may rewrite Eq.~\eqref{eq:CircleSimple0} as 
\begin{equation}
    \label{eq:CircleConicImplicit}
    \bar{\bp}^T_L \bC_L \bar{\bp}_L = 0
\end{equation}
where $\bar{\bp}_L \in \PP^2$ is given by
\begin{equation}
    \bar{\bp}_L \propto \begin{bmatrix}
        X \\ Y \\ 1
   \end{bmatrix}
\end{equation}

Now suppose this CoL is viewed by a camera that produces images obeying perspective projection. 
Following the camera frame conventions of Ref.~\cite{Christian:2021d} that places the camera's $+z$-axis out of the camera, a homogeneous point $\bar{\bp}$ lying on the circle appears in a 2-D image at homogeneous coordinates $\bar{\bx} \in \PP^2$
\begin{equation}
    \begin{bmatrix}
        x \\ y \\ 1
   \end{bmatrix} = \bar{\bx} \propto \bT^L_C \bS \bar{\bp}_L + \br_C
\end{equation}
where $\br_C \in \RR^3$ is the vector from the camera to the center of the CoL (expressed in the camera frame $C$), $\bT^L_C = [\bt_1, \bt_2, \bt_3]$ is the rotation matrix from frame $L$ to $C$ with columns given by $\{\bt_i\}_{i=1}^3$, and $\bS$ is the matrix
\begin{equation}
    \bS =  \begin{bmatrix}
        1 & 0 & 0\\ 0 & 1 & 0 \\ 0 & 0 & 0
   \end{bmatrix}
\end{equation} 
Which, simplifies to
\begin{equation}
   \bar{\bx} \propto \bH \bar{\bp}
\end{equation}
where $\bH$ is the homography matrix
\begin{equation}
   \bH \propto \begin{bmatrix}
        \bt_1 & \bt_2 & \br_C
   \end{bmatrix}
\end{equation}
and it follows that
\begin{equation}
    \bar{\bp} \propto \bH^{-1} \bar{\bx}
\end{equation}
Substituting this result into Eq.~\eqref{eq:CircleConicImplicit},
\begin{equation}
    \bar{\bp}^T \bC \bar{\bp} = \bar{\bx}^T\bH^{-T} \bC \bH^{-1} \bar{\bx} = 0
\end{equation}
Thus, defining $\bA$ to be the symmetric $3 \times 3$ matrix of arbitrary scale defining the conic resulting from projection of the CoL into the image plane,
\begin{equation}
    \bA \propto \bH^{-T} \bC \bH^{-1}
\end{equation}
and one observes an image plane conic  of
\begin{equation}
    \label{eq:ImagePlaneConicQuad}
    \bar{\bx}^T \bA \bar{\bx} = 0
\end{equation}
Defining the elements of $\bA$ to be,
\begin{equation}
    \label{eq:DefA}
    \bA = \begin{bmatrix}
        A & B/2 & D/2 \\
        B/2 & C & F/2 \\
        D/2 & F/2 & G
    \end{bmatrix}
\end{equation}
then it follows that Eq.~\eqref{eq:ImagePlaneConicQuad} describes the implicit equation for a conic
\begin{equation}
    \label{eq:ImplicitConic}
    A x^2 + Bxy + Cy^2 + Dx + Fy + G = 0
\end{equation}
which is an ellipse (or circle) when $B^2 < 4AC$. Equations relating the coefficients $A,\hdots,G$ to geometric parameters (e.g., ellipse center, semi-major axis) are given in the appendix. This will become important in Section~\ref{Sec:PoleCovariance} for the development of analytic covariance expressions.

Points on the image plane $[x,y]$ are related to the pixel coordinates $[u,v]$ in the corresponding digital image through an affine transformation,
\begin{equation}
    \begin{bmatrix}
        u \\ v \\ 1
   \end{bmatrix} = \bar{\bu} =  \bK \bar{\bx}
\end{equation}
where $\bK$ is the so-called camera calibration matrix (see Refs.~\cite{Hartley:2003} and \cite{Christian:2021d}). In almost all space applications the camera is calibrated and the matrix $\bK$ is known (e.g., Refs.~\cite{Christian:2016} and \cite{Golish:2020}). Thus, a measured pixel coordinate $\bar{\bu}$ may always be transformed into the image plane according to $\bar{\bx} = \bK^{-1} \bar{\bu}$. Moreover, suppose an ellipse is fit to observed image points $[u,v]$ using a standard ellipse fitting method \cite{Fitzgibbon:1999,Chojnacki:2000,AlSharadqah:2009,Kanatani:2011,Szpak:2015} (with the ``semihyper'' least squares method from Ref.~\cite{Kanatani:2011} being especially recommended). If the image ellipse in pixel coordinates is described by the matrix $\bB$ such that,
\begin{equation}
   \bar{\bu}^T \bB \bar{\bu} = 0
\end{equation}
then the camera calibration matrix may be used to transform image ellipse $\bB$ to image plane ellipse $\bA$ according to
\begin{equation}
   \bA \propto \bK^T \bB \bK
\end{equation}
Consequently, the remainder of this work operates entirely in image plane coordinates $\bar{\bx}$ and image plane ellipses $\bA$---both of which may be readily obtained from a digital image collected by a calibrated camera.

\section{Estimating Pole Direction from Circle of Latitude Projections}
\label{Sec:PoleDirection}
The orientation of the pole for a rotating celestial body coincides with the normal direction to the plane of every CoL. Since each CoL projects to an ellipse in an image, the apparent image ellipse may be used to determine the pole orientation as expressed in the camera frame. This may be achieved by realizing that an image ellipse back-projects to a cone. If the observed image ellipse is known to originate from a circular feature (e.g., a CoL), then the objective is to find the two plane orientations whose intersection with this cone create a circle. This is a classical and well-studied result. While the solutions to this problem vary slightly \cite{Shiu:1989,Dhome:1990,Forsyth:1991,Kanatani:1993,Christian:2017AAS}, most fundamentally follow a two step procedure: (1) back-project the image ellipse to get a cone and (2) find the two plane orientations that cut a circle in this cone. These two steps are now discussed, followed by the development of an analytic covariance model.

\subsection{Cone Back-Projection}
The back-projection of rays from the camera through the image ellipse forms a conical surface \cite{Hartley:2003}. To see this, define a 3-D point $\by\in \RR^3$ lying on the ray described by image plane point $\bar{\bx}$ such that $\by = a \bar{\bx}$ for $a > 0$. Substituting this into Eq.~\eqref{eq:ImagePlaneConicQuad}, any point $\by$ belonging to the cone must satisfy
\begin{equation}
    \by^T \bA \by = 0
\end{equation}
The observed image plane ellipse described by $\bA$ is the conic section of this cone formed by an intersection with the camera frame's $z=1$ plane. Likewise, in the present application, the circle of latitude is a different conic section of this same cone.

Since $\bA$ is a symmetric $3 \times 3$ matrix, an eigendecomposition may be used to diagonalize the system. Moreover, since $\bA$ is guaranteed to be an indefinite matrix when describing an ellipse, the result is three eigenvalues $\{\lambda_i\}^3_{i=1}$ of mixed sign. Thus, recalling that the scale (and sign) of $\bA$ is arbitrary, one can write without loss of generality
\begin{equation}
    \label{eq:ConeEigenvalueOrder}
    \lambda_1 \geq \lambda_2 > 0 > \lambda_3
\end{equation}
which will always be the case if $\bA$ is scaled such that $\text{det}(\bA)=-1$. This is the particular scaling of $\bA$ used in all numerical examples within this work, though different scalings are equally reasonable (so long as one is careful with the eigenvalue sign conventions).   

The three corresponding unit eigenvectors $\{\bu_i\}_{i=1}^3$ form an orthonormal basis (since $\bA$ is symmetric) and describe the principal axes of the cone (as thoroughly discussed in Ref.~\cite{Christian:2017AAS}). This is denoted as frame $P$. The eigenvector corresponding to the eigenvalue of unique sign describes the cone's centerline, which will be $\bu_3$ given the convention from Eq.~\eqref{eq:ConeEigenvalueOrder}.
Moreover, since the signs of $\pm \bu_i$ are ambiguous, choose the sign of $\pm \bu_3$ to be positive out of the camera: $\bk^T \bu_3 > 0$, where $\bk^T = [0, 0, 1]$. Also make the system right-handed by adjusting the sign of $\pm \bu_2$ to ensure that $\bu_1 \times \bu_2 = \bu_3$. Therefore, one may construct a rotation matrix to transform vectors from the cone principal axis frame $P$ to the camera frame $C$ as
\begin{equation}
    \label{eq:RotationConeToCamera}
    \bT^P_C = \begin{bmatrix}
        \bu_1 & \bu_2 & \bu_3
    \end{bmatrix}
\end{equation}

\subsection{Estimating Pole Direction from a Single Circle Projection}
Given a cone, it is always possible to find two plane orientations whose conic sections are a circle. Different planes of this orientation will produce circles of different radius, which will become important later. There are a few different ways to describe these two planes found in the literature and they will be briefly reviewed here (and shown to be equivalent). 

It has been shown by numerous authors \cite{Shiu:1989,Christian:2017AAS} that the two normal directions $\bn_P \in \PP^2$ producing circular conic sections in the cone principal axis frame are given by
\begin{equation}
    \label{eq:PlaneNormalCone0}
    \bn_{P} \propto 
    \begin{bmatrix}
        \pm \sqrt{\alpha} \\ 0 \\ -\sqrt{ -\lambda_3 / \lambda_1}
    \end{bmatrix}
\end{equation}
where
\begin{equation}
    \label{eq:DefAlpha}
    \alpha = \frac{1/\lambda_2 - 1/\lambda_1}{1/\lambda_2 - 1/\lambda_3} = \frac{\lambda_3}{\lambda_1}\frac{\lambda_1 - \lambda_2}{\lambda_3 - \lambda_2}
\end{equation}
Note the proportional sign in Eq.~\eqref{eq:PlaneNormalCone0} which indicates that $\bn_P \in \PP^2$ is not a unit vector and simply defines a direction in $\RR^3$. Since the physical interpretation of this can be somewhat opaque, substitute for $\alpha$ in Eq.~\eqref{eq:PlaneNormalCone0} and expand
\begin{equation}
    \label{eq:NormalInConeFrameSimple}
    \bn_{P} \propto 
    \begin{bmatrix}
        \pm \sqrt{(\lambda_1 - \lambda_2)/(\lambda_2 - \lambda_3)} \\ 0 \\ 1
    \end{bmatrix} \propto
    \begin{bmatrix}
        \pm \sqrt{\lambda_1 - \lambda_2} \\ 0 \\ \sqrt{\lambda_2 - \lambda_3}
    \end{bmatrix} 
\end{equation}
which represents a rotation of the vector $\bk = [0,0,1]^T$ about the $P$ frame's $y$-axis by the angle $\pm \theta$,
\begin{equation}
    \label{eq:CirclePlaneAngle}
   \tan \pm \theta = \sqrt{ \frac{\lambda_1 - \lambda_2}{\lambda_2 - \lambda_3} }
\end{equation}
which is exactly the angle described by Forsyth, et al. \cite{Forsyth:1991}. Thus, one could equivaletnly write $\bn_P \propto \bT_2(\pm \theta) \, \bk$, where $\bT_2( \, \cdot \, )$ is the elementary rotation matrix about the second coordinate direction (i.e., $y_P$-axis).

It is sometimes desirable to write $\bn_P$ explicitly as the unit vector $\hat{\bn}_P = \bn_P / \| \bn_P \|$. Therefore, observe that the right-hand side of Eq.~\eqref{eq:NormalInConeFrameSimple} has a norm of 
\begin{equation}
    \sqrt{ (\lambda_1 - \lambda_2) + (\lambda_2 - \lambda_3) } = \sqrt{ \lambda_1 - \lambda_3 }
\end{equation}
and so the unit normal in the cone's principal axis frame becomes
\begin{equation}
    \label{eq:UnitNormalInConeFrame}
    \hat{\bn}_{P} =
    \begin{bmatrix}
        \pm \sqrt{(\lambda_1 - \lambda_2)/(\lambda_1 - \lambda_3 )} \\ 0 \\ \sqrt{(\lambda_2 - \lambda_3)/(\lambda_1 - \lambda_3 )}
    \end{bmatrix} 
\end{equation}
Finally, since the rotation back into the camera frame is given by $\bT^P_C$ from Eq.~\eqref{eq:RotationConeToCamera}
\begin{equation}
    \bn_C \propto \bT^P_C \bn_P
\end{equation}
which may be written out as as a linear combination of the eigenvectors
\begin{equation}
    \bn_C \propto \sqrt{\lambda_1 - \lambda_2} \; \bu_1 \pm \sqrt{\lambda_2 - \lambda_3} \; \bu_3
\end{equation}
or, if desired, explicitly as a unit vector
\begin{equation}
    \label{eq:UnitNormalFinal}
    \hat{\bn}_C =  \sqrt{\frac{\lambda_1 - \lambda_2}{\lambda_1 - \lambda_3 }} \; \bu_1 \pm \sqrt{\frac{\lambda_2 - \lambda_3}{\lambda_1 - \lambda_3 }} \; \bu_3
\end{equation}
which agrees with the result presented in \cite{Kanatani:1993}. Note that the eigenvalue convention from Eq.~\eqref{eq:ConeEigenvalueOrder} guarantees that $\lambda_1 - \lambda_3 > 0$ and there is no risk of dividing by zero.  In summary, the different formulations of Refs.~\cite{Shiu:1989,Dhome:1990,Forsyth:1991,Kanatani:1993,Christian:2017AAS} all give the exact same result and there is no compelling reason to prefer one over the other. The reader interested in a detailed derivation is directed to these references.

It is insightful to consider what happens in the case of orthographic projection---which one approaches in the limit when the observed body is along the camera boresight and the distance becomes very large. Images from narrow field of view cameras or telescopes often appear nearly orthographic, and so this limiting case helps build geometric intuition. The reader is strongly cautioned, however, to not use this limiting geometry for pole estimation as its performance does not deteriorate gracefully. The analytic result from Eq.~\eqref{eq:UnitNormalFinal} should always be used to actually solve for the pole.  Therefore, considering the limiting case of orthographic projection, observe that
\begin{equation}
    \sqrt{\frac{\lambda_1 - \lambda_2}{\lambda_1 - \lambda_3 }} \; \rightarrow \; \sqrt{1 - b^2/a^2} = e
\end{equation}
\begin{equation}
    \sqrt{\frac{\lambda_2 - \lambda_3}{\lambda_1 - \lambda_3 }} \; \rightarrow \;  b/a = \epsilon
\end{equation}
where $e$ is the ellipse eccentricity, $\epsilon$ is the ellipse ellipticity, and $a$ and $b$ are the image ellipse semi-major and semi-minor distances.
Recalling the observation from Ref.~\cite{Forsyth:1991} discussed in Eq.~\eqref{eq:CirclePlaneAngle}, if the angle $\theta$ is the angle between the cone principal axis $\bu_3$ and the normal vector of the circular feature $\hat{\bn}_C$, then $\sin \theta \rightarrow e$ and $\cos \theta \rightarrow \epsilon$ as one approaches orthographic projection and it is readily apparent that
\begin{equation}
    1 = \sin^2 \theta + \cos^2 \theta  \rightarrow e^2 + \epsilon^2 = 1
\end{equation}
It follows, then, that the unit normal $\hat{\bn}_C$ tends towards
\begin{equation}
    \label{eq:NormalLimit0}
    \hat{\bn}_C \; \rightarrow \; e \; \bu_1 \pm \epsilon \; \bu_3
\end{equation}
which remains exactly a unit vector, even if $e$ and $\epsilon$ do not exactly correspond to the correct normal orientation $\theta$. Again, while helpful for geometric intuition, the reader is urged to not use Eq.~\eqref{eq:NormalLimit0} for estimating $\hat{\bn}_C$. %, but is only helpful within the context of developing an approximate covaraince expression. 
The approximation of Eq.~\eqref{eq:NormalLimit0} does not appear especially robust as the geometry departs from orthographic projection. The practical computation of $\hat{\bn}_C$ should always be accomplished with Eq.~\eqref{eq:UnitNormalFinal} or one of it's equivalents.

\subsection{Approximate Covariance of Pole Direction for a Single Circle Projection}
\label{Sec:PoleCovariance}
It is often desirable to know the error covariance for the estimated unit vector $\hat{\bn}_C$ from Eq.~\eqref{eq:UnitNormalFinal}. This task is made more difficult than usual since Eq.~\eqref{eq:UnitNormalFinal} depends on the eigenvectors and eigenvalues of $\bA$ which change in a complicated manner with perturbations in the image ellipse's apparent shape. Fortunately, however, the error covariance of $\hat{\bn}_C$ may still be found analytically using well-understood procedures for computing the Jacobian of eigenvalues and eigenvectors \cite{Magnus:1985,Liounis:2019}. The methods developed here work well everywhere except when camera's position is nearly along the 3-D line formed by the observed body's pole. Such a vantage point leads to the condition $\lambda_1 \approx \lambda_2$ where Eq.~\eqref{eq:UnitNormalFinal} is not differentiable. Although the estimate of $\hat{\bn}_C$ from Eq.~\eqref{eq:UnitNormalFinal} is still valid, a different means of developing the covariance in this special scenario is needed---and this task is left to future work. Therefore, assuming the camera is not perfectly along the pole (i.e., that $\lambda_1 \neq \lambda_2$), the development is briefly summarized here and the full partial derivatives are provided in the appendix. 

Defining $\blam^T = [\lambda_1, \lambda_2, \lambda_3]$ to be a vector of the three eigenvalues of $\bA$, compute the $3 \times 9$  Jacobian matrix $\bbN$ as 
\begin{equation}
    \label{eq:PartialsN}
    \bbN = \begin{bmatrix} \partial \hat{\bn}_C / \partial \blam & \partial \hat{\bn}_C / \partial \bu_1 & \partial \hat{\bn}_C / \partial \bu_3 \end{bmatrix}
\end{equation}
Then compute the $9 \times 9$ Jacobian of the eigenvalues and relevant eigenvectors with respect to the vectorized matrix $\bA$
\begin{equation}
    \label{eq:PartialsM}
    \bM = \begin{bmatrix} \partial \blam / \partial \text{vec}(\bA) \\
    \partial \bu_1 / \partial \text{vec}(\bA) \\
    \partial \bu_3 / \partial \text{vec}(\bA) 
     \end{bmatrix}
\end{equation}
Now, since $\bA$ is symmetric [see Eq.~\eqref{eq:DefA}], define the $9 \times 6$ matrix $\bP$ (see appendix) such that $\text{vec}(\bA) = \bP \ba$ and where $\ba = [A,B,C,D,F,G]^T$ is a $6 \times 1$ vector of the coefficients of the implicit equation for an ellipse [see Eq.~\eqref{eq:ImplicitConic}]. Consequently, one may write the Jacobian of interest as
\begin{equation}
    \partial \hat{\bn}_C / \partial \ba = \bbN \bM \bP
\end{equation}
such that
\begin{equation}
    \delta \hat{\bn}_C = \bbN \bM \bP \delta \ba
\end{equation}
Thus, the covariance of the unit pole direction $\hat{\bn}_C$ is
\begin{equation}
    \label{eq:PoleCovariance}
    \bR_{\hat{\bn}_C} = E[\delta \hat{\bn}_C \, \delta \hat{\bn}^T_C] = \bbN \bM \bP \bR_{\ba} \bP^T \bM^T \bbN^T
\end{equation}
where $\bR_a = E[\delta \hat{\ba} \, \delta \hat{\ba}^T]$ is the known error covariance of the ellipse fit. The covariance $\bR_a$ is a $6 \times 6$ matrix with $\text{rank}[\bR_a]=5$ since $\ba$ has arbitrary scale.  Note that this covariance was intentionally developed in terms of errors in $\ba$ because most modern ellipse fitting algorithms estimate $\ba$ directly (and not things like ellipse center location, semi-major axis, and so on), and so statistics on an ellipse fit are most readily available in terms of $\ba$. If statistics are instead known for the five geometric parameters $\bg = [a,b,x_c,y_c,\phi]^T$, then one may compute $\bR_a$ as
\begin{equation}
    \bR_{\ba} = \left( \frac{\partial \ba}{\partial \bg} \right) \bR_\bg \left( \frac{\partial \ba}{\partial \bg} \right)^T 
\end{equation}
where $\bR_\bg = E[\delta \hat{\bg} \, \delta \hat{\bg}^T]$ is the error covariance of $\bg$.

\subsection{Estimating the Pole from Many Circle Projections}
Each CoL produces a pair of possible solutions for $\hat{\bn}_C$ from Eq.~\eqref{eq:UnitNormalFinal}. Except when looking nearly down the pole, these two directions are distinct. Thus, if one has many CoL projections, it is simple to unambiguously separate the estimates of $\hat{\bn}_C$ into two groups. In the idealized (and unrealistic) case of perfect CoL projections with no measurement noise, one group of solutions for $\hat{\bn}_C$ will be exactly the same for every projection and the other group of solutions will be slightly different from one another. However, with realistic measurement noise, the differences in the incorrect choice of $\hat{\bn}_C$ are so small that they aren't useful in distinguishing which of the two choices for $\hat{\bn}_C$ is correct. Thus, without additional contextual information, it is often difficult to determine which of the two pole orientation estimates corresponds to the true pole orientation. Usually, however, a simple inspection of the image or an occlusion check will allow the analyst to select the correct pole solution. 

Proceeding under the assumption that the pole ambiguity has been resolved, it is possible to construct a maximum likelihood estimate (MLE) of the pole orientation from many CoL projections. Since an analytic pole covariance is available for each individual pole estimate, the MLE solution takes the form of a weighted least squares problem \cite{Tapley:2004}. The solution to this is, therefore, given by
\begin{equation}
    \label{eq:PoleMLE}
    \hat{\bn}_C = \left[ \sum_{i=1}^n \bR^{-1}_{\hat{\bn}_{C,i}} \right]^{-1} \sum_{i=1}^n \bR^{-1}_{\hat{\bn}_{C,i}} \hat{\bn}_{C,i}
\end{equation}
with covariance
\begin{equation}
    \label{eq:PoleMLEcovariance}
    \bR = \left[ \sum_{i=1}^n \bR^{-1}_{\hat{\bn}_{C,i}} \right]^{-1}
\end{equation}
When no covariance is available (e.g, when no estimate of ellipse fit quality is available) then each observation is weighted equally and Eq.~\eqref{eq:PoleMLE} collapses to a simple average
\begin{equation}
    \hat{\bn}_C = \frac{1}{n} \sum_{i=1}^n \hat{\bn}_{C,i}
\end{equation}

\section{Recovering Three-Dimensional Circle of Latitude Structure}
When more than one CoL projection is observed, it is possible to infer more information about the observed body than just the pole orientation.
If $\hat{\bn}_{C}$ is as shown in Eq.~\eqref{eq:UnitNormalFinal}, then it was shown in Refs.\cite{Shiu:1989} and \cite{Christian:2017AAS} that vector from the camera to the circle center is
\begin{equation}
    \br_{C} = \frac{R}{\sqrt{\alpha - \lambda_3 / \lambda_1}} \bT^P_C
    \begin{bmatrix}
        \mp \sqrt{-\alpha \lambda_3 / \lambda_1} \\ 0 \\ 1
    \end{bmatrix}
\end{equation}
where $R$ is the CoL radius, $\bT^P_C$ is from Eq.~\eqref{eq:RotationConeToCamera}, and $\alpha$ is from Eq.~\eqref{eq:DefAlpha}.
If the radius $R$ is not (yet) known, then one may estimate the relative position as normalized by the circle radius. Therefore, defining $\brho_{C} = \br_{C} / R$, it is always possible compute $\brho_{C}$ as
\begin{equation}
    \label{eq:DefRhoC}
    \brho_{C} = \frac{\br_{C}}{R}= \frac{1}{\sqrt{\alpha - \lambda_3 / \lambda_1}} \bT^P_C
    \begin{bmatrix}
        \mp \sqrt{-\alpha \lambda_3 / \lambda_1} \\ 0 \\ 1
    \end{bmatrix}
\end{equation}
where everything on the right-hand side is known from the eigenvectors and eigenvalues of the ellipse shape matrix $\bA$. This result may be used to determine the 3-D structure to an unknown global scale.

\subsection{Scaled Three-Dimensional Structure}
\label{Sec:Scaled3DStructure}
Suppose there are $n \geq 2$ image ellipses formed by the projection of $n$ unknown CoLs. Denote the circle radii as $\{R_i\}_{i=1}^n$ and the circle relative positions as $\{ \br_{C_i} \}_{i=1}^n$. If the CoLs all come from the same spinning celestial object then they must have the same normal, hence $\hat{\bn}_{C_i} = \hat{\bn}_C$ for $i=1,\hdots,n$. Choose one of the elliptical projections as the reference feature and define the corresponding CoL's radius and relative position as $R_r$ and $\br_{C_r}$.

\begin{figure}[b!]
    	\centering
    	\includegraphics[width=0.7\textwidth]{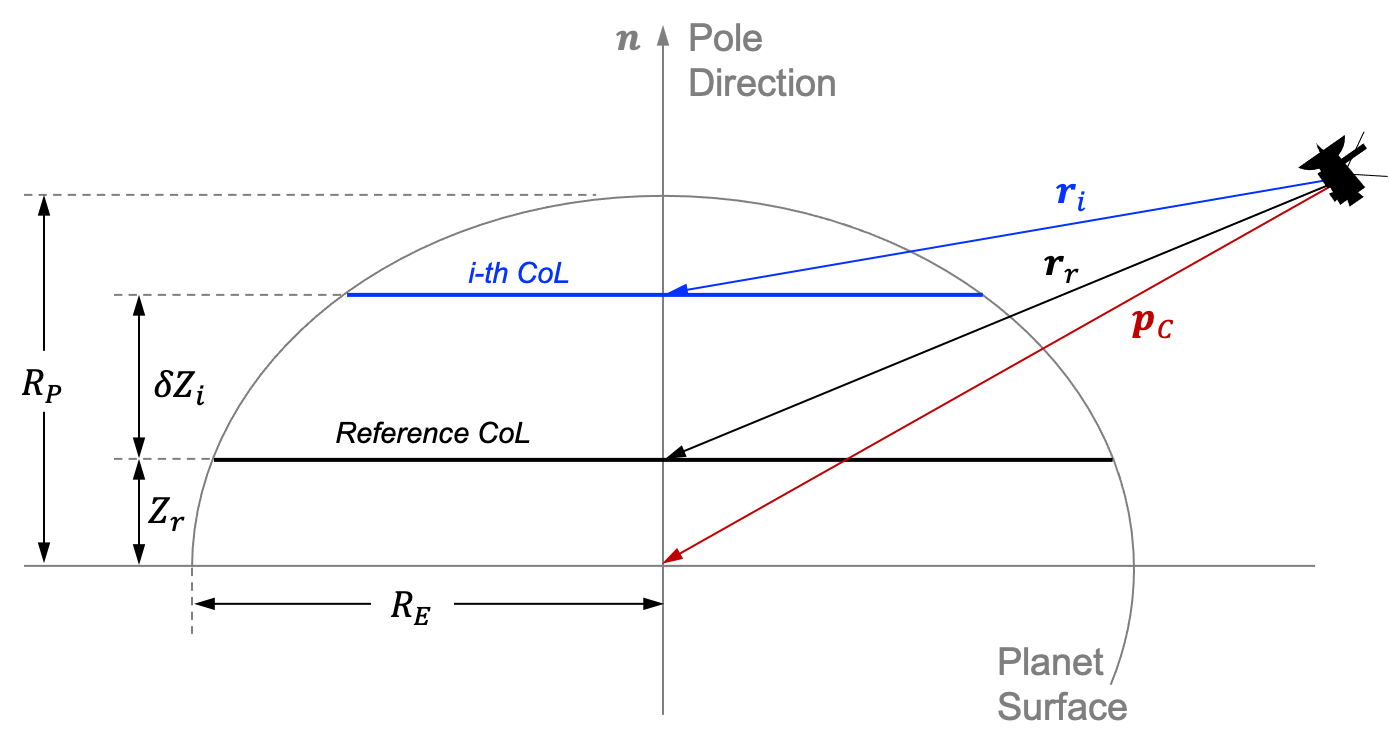}
    	\caption{Illustration of multiple CoL geometry that may be used for 3-D reconstriction.}
    	\label{fig:SpheroidCoL}
\end{figure}

Since every CoL center lies along the pole, they may all be written as
\begin{equation}
    \label{eq:RelativePositionWithScale}
    \br_{C_i} = \br_{C_r} + \delta Z_{i} \hat{\bn}_C
\end{equation}
where $\delta Z_i = Z_i - Z_r$ is the distance (positive in the direction of $\bn_C$) from the reference CoL to the $i$-th CoL. This geometry is illustrated in Fig.~\ref{fig:SpheroidCoL} (where frame subscripts have been removed). 

However, since the relative positions $\br_{C_r}$ and $\br_{C_i}$ may not be directly obtained from the image,
one may instead rewrite Eq.~\eqref{eq:RelativePositionWithScale} as
\begin{equation}
    R_i \brho_{C_i} = R_r \brho_{C_r} + \delta Z_{i} \hat{\bn}_C 
\end{equation}
where $\hat{\bn}_C$, $\brho_{C_r}$, and $\brho_{C_i}$ are all known from the methods above [see Eqs.~\eqref{eq:UnitNormalFinal} and \eqref{eq:DefRhoC}]. Clearly, $R_r$, $R_i$, and $\delta Z_{i}$ may not be simultaneously determined in an absolute sense. Therefore, normalize by $R_r$ to obtain
\begin{equation}
   (R_i/R_r) \brho_{C_i} = \brho_{C_r} + (\delta Z_{i}/R_r) \hat{\bn}_C 
\end{equation}
Letting the normalized distances $R'_i$ and $\delta Z'_{i}$ be 
\begin{subequations}
\label{Eq:ScaledCoL}
\begin{equation}
    R'_i = R_i/R_r
\end{equation}
\begin{equation}
    \delta Z'_{i} = \delta Z_{i}/R_r
\end{equation}
\end{subequations}
one obtains the $3 \times 2$ linear system in these two unknowns
\begin{equation}
    \label{eq:LinSysReconstruct}
   \brho_{C_r} = \begin{bmatrix} \brho_{C_i} & -\hat{\bn}_C  \end{bmatrix} 
   \begin{bmatrix} R'_i \\ \delta Z'_{i}  \end{bmatrix}
\end{equation}
This system may be directly solved for $R'_i$ and $\delta Z'_{i}$. In the case of the reference CoL itself (i.e., when $i=r$), this produces the expected result of $R'_r = 1$ and $\delta Z'_r = 0$. Observability into $R'_i $ and $\delta Z'_i$ disappears when $\brho_{C_i}$ and $\hat{\bn}_C$ are parallel (or nearly parallel), which occurs when the camera lies on the line defining the poles (i.e., is located directly above/below the poles). This happens to be the same degenerate configuration in which Eq.~\eqref{eq:UnitNormalFinal} for the normal direction $\bn_C$ is not differentiable, leading to difficulties with computing the pole direction covariance. Regardless, outside of this single degenerate viewpoint, the result of Eq.~\eqref{eq:LinSysReconstruct} may be used to reconstruct the 3-D system of CoLs to an unknown global scale---with this unknown scale being the radius $R_r$ of the reference CoL. 
The unknown global scale may be resolved in some special cases, as discussed in Section~\ref{Sec:ScaleForSpheroid}.

\subsection{Three-Dimensional Pole Representations with Pl\"{u}cker Coordinates}
It may sometimes be useful to describe the 3-D line in space describing the pole of the spinning body. A 3-D line may be represented as a point in $\PP^5$ and may often be efficiently described with Pl\"{u}cker coordinates \cite{Plucker:1865,Hartley:2003}. The location of the center of circle $\br_{C_r}$ and the pole direction $\bn_C$ may be combined to describe a 3-D line in the camera frame. Specifically, following Ref.~\cite{Christian:2021streak}, one finds that
\begin{equation}
   \label{eq:PluckerMat}
    \bL_C \propto \begin{bmatrix}
        \left[(\bn_C \times \br_C) \times \right] & -\bn_C\\
        \bn_C^T & 0
    \end{bmatrix}
\end{equation}
which projects to the line in the image according to
\begin{equation}
   \bl \propto \bn_C \times \br_C 
\end{equation}
where image points belonging to the projected 2-D line $\bl$ satisfy the constraint $\bl^T \bar{\bx} = 0$. The 2-D image line $\bl$ passes through the projected points corresponding to the centers of each of the 3-D circles (which are points along the pole). Note, however, that the line does \emph{not} necessarily pass through the center of the projected ellipses since the circle centers to not project to the ellipse centers. In the event that scale is not known, $\brho_C$ may be used in place of $\br_C$ in Eq.~\eqref{eq:PluckerMat} to obtain the line for the scaled 3-D geometry.

\section{Camera Localization for an Oblate Spheroid}
\label{Sec:ScaleForSpheroid}

Most of the large celestial bodies in our Solar System may be well-modeled as an oblate spheroid. This occurs since rotating and self-gravitating bodies in hydrostatic equilibrium (e.g., Jupiter) take the shape of an oblate spheroid, as was shown by Newton in the \emph{Principia} \cite{Newton:1687}. Thus, since many of the large celestial bodies in our Solar System have a well-known spheroidal shape, the scale-ambiguous CoL structure from Section~\ref{Sec:Scaled3DStructure} may be adjusted to tightly lie on the spheroidal surface.

If the celestial body is an oblate spheroid, then its shape is defined by the known equatorial radius $R_E$ and polar radius $R_P$. Thus, any CoL must satisfy the constraint
\begin{equation}
   \frac{R^2_i}{R_E^2} + \frac{Z_i^2}{R_P^2} = 1
\end{equation}
which is the same as
\begin{equation}
   \frac{R^2_i}{R_E^2} + \frac{(Z_r + \delta Z_i)^2}{R_P^2} = 1
\end{equation}
Proceed by substituting from Eq.~\eqref{Eq:ScaledCoL}
\begin{equation}
   \frac{R^2_r R'^2_i}{R_E^2} + \frac{R^2_r(Z'_r + \delta Z'_i)^2}{R_P^2} = 1
\end{equation}
which, defining $\epsilon = R_P/R_E \leq 1$, may be rearranged to find
\begin{equation}
   (\epsilon^2 R'^2_i + \delta Z'^2_i) R^2_r + (2 \delta Z'^2_i) R^2_r Z'_r + (R_r^2 Z'^2_r - R_P^2) = 0
\end{equation}
Here, it is noted that $\epsilon$ is known since $R_E$ and $R_P$ are known. Finally, rearranging leads to the linear system
\begin{equation}
   \begin{bmatrix}
       (\epsilon^2 R'^2_1 + \delta Z'^2_1)& (2 \delta Z'^2_1) & 1 \\
       \vdots & \vdots & \vdots \\
       (\epsilon^2 R'^2_n + \delta Z'^2_n)& (2 \delta Z'^2_n) & 1
   \end{bmatrix}
   \bxi = 
   \begin{bmatrix}
       0 \\
       \vdots  \\
       0
   \end{bmatrix}
\end{equation}
\begin{equation}
    \bxi = \begin{bmatrix}
        \xi_1 \\ \xi_2 \\ \xi_3
   \end{bmatrix}  \propto
   \begin{bmatrix}
        R^2_r \\ R^2_r Z'_r \\ R_r^2 Z'^2_r - R_P^2
   \end{bmatrix} 
\end{equation}
This is a null space problem, so all that is important are the ratios
$\xi_1:\xi_2:\xi_3$.
The optimal solution for $\bxi$ is found via the singular value decomposition. With an estimate of $\bxi$ in hand,
one can immediately recover the scaled offset as $Z'_r = \xi_2 / \xi_1$. Likewise, one can recover the reference CoL radius $R_r$ as
\begin{equation}
    \label{Eq:RefCoLRadiusSpheroid}
    R_r = R_P \, \sqrt{ \frac{ \xi_1^2}{\xi_2^2 - \xi_1 \xi_3}  }
\end{equation}
Therefore, since $Z_r = R_r Z'_r$, one finds that
\begin{equation}
    \label{Eq:RefCoLOffsetSpheroid}
    Z_r = R_P \, \sqrt{ \frac{ \xi_2^2}{\xi_2^2 - \xi_1 \xi_3}  }
\end{equation}
Consequently, from the geometry in Fig.~\ref{fig:SpheroidCoL}, the position from the spacecraft to the center of the oblate spheroid as expressed in the camera frame
\begin{equation}
    \bp_C = R_r \brho_{C_r} - Z_r \hat{\bn}_C
\end{equation}
where $\hat{\bn}_C$ is from Eq.~\eqref{eq:PoleMLE}, $\brho_{C_r}$ is from Eq.~\eqref{eq:DefRhoC}, $R_r$ is from Eq.~\eqref{Eq:RefCoLRadiusSpheroid}, and $Z_r$ is from Eq.~\eqref{Eq:RefCoLOffsetSpheroid}.

\section{Numerical Results}

\subsection{Example with a Spinning Asteroid}
\label{Section:BennuExample}
\begin{figure}[t!]
    	\centering
    	\includegraphics[width=0.5\textwidth]{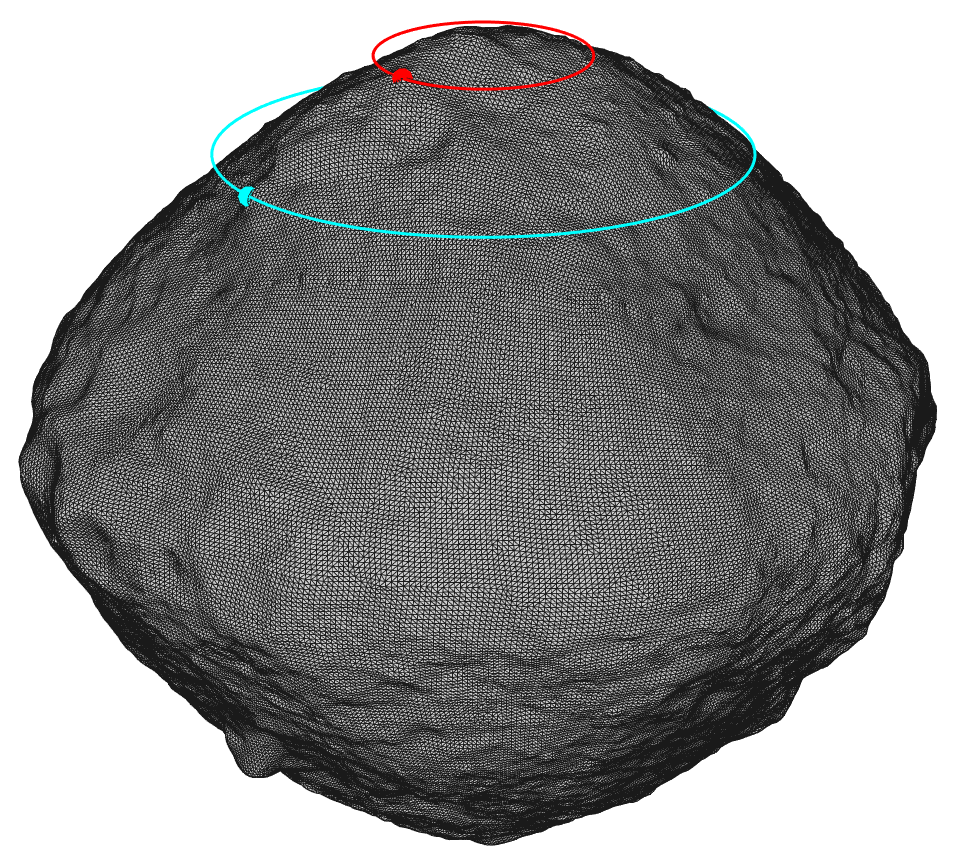}
    	\caption{Mesh model of Bennu with two CoLs (red and cyan circles) produced by two reference surface points (red and cyan dots).}
    	\label{fig:BennuCoL}
\end{figure}

As the first numerical example, consider the scenario where the motion of surface points on a spinning asteroid create elliptical paths in a sequence of images. The asteroid Bennu is used as the reference model to allow for easy comparison with the prior work of Kuppa, McMahon, and Dietrich \cite{Kuppa:2022a,Kuppa:2022b}. Suppose that ellipses are fit to the projections of two CoLs. Each of these two CoLs pass through the surface point that generated them. However, since Bennu is not a body of revolution (e.g., an oblate spheroid), these CoLs do not necessarily lie on any other point on Bennu's surface---this can be seen in Fig.~\ref{fig:BennuCoL}. 

\begin{figure}[t!]
    	\centering
    	\includegraphics[width=0.5\textwidth]{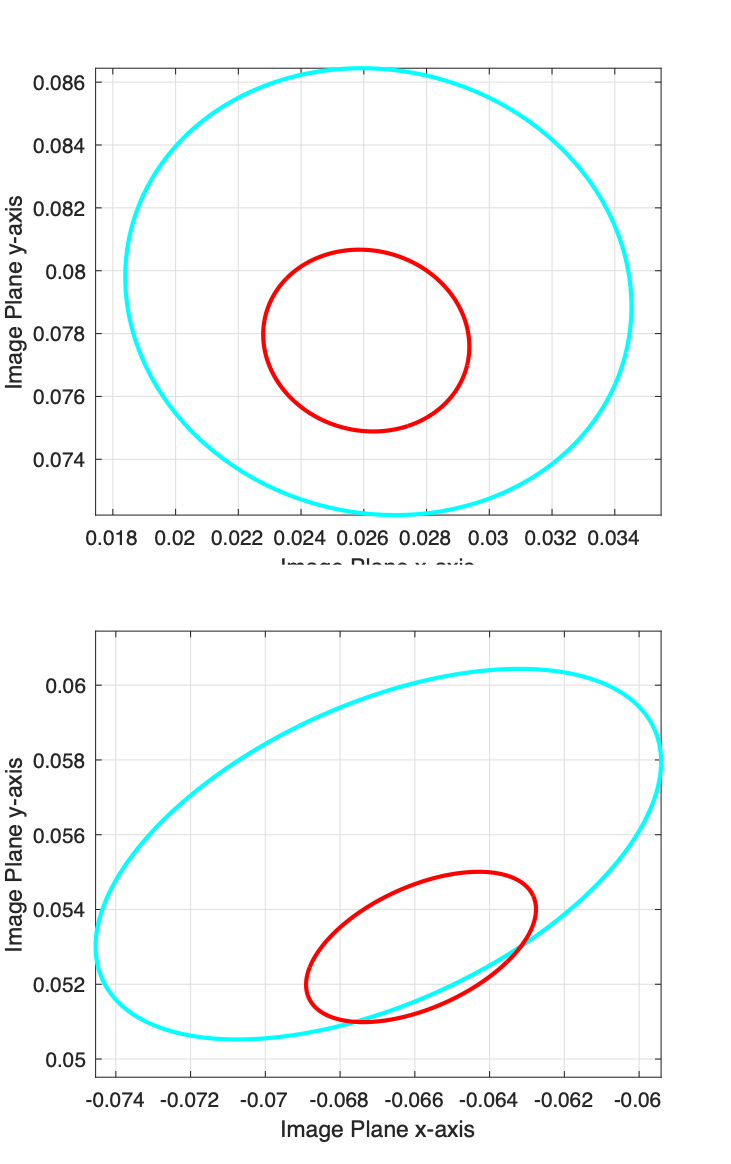}
    	\caption{Projection of CoLs from Fig.~\ref{fig:BennuCoL} onto the image plane for a spacecraft at a latitude of 60 degrees (top) and 30 degrees (bottom).}
    	\label{fig:BennuProjEllipse}
\end{figure}

Within this scenario, two specific configurations are considered: (1) a spacecraft at a latitude of 60 deg and (2) a spacecraft at a latitude of 30 deg. In both configurations, the camera is not pointed directly at the center of Bennu or at the center of either of the observed CoLs. These projections produce the image plane ellipses shown in Fig.~\ref{fig:BennuProjEllipse}. To process these image ellipses, the algorithm is not provided any information about the camera-relative position or orientation of Bennu---only the image-plane ellipse fits [as parameterized by the matrix $\bA$ from Eq.~\eqref{eq:DefA}] are available. Note that the method from Ref.~\cite{Kuppa:2022a,Kuppa:2022b} does not work in such a situation, since it requires the camera to be pointed at the asteroid center and that the distance is known.

Begin by considering the noise free case. The pole directions are computed using Eq.~\eqref{eq:UnitNormalFinal} and are shown in Tables~\ref{tab:AsteroidPerfectPole60} and \ref{tab:AsteroidPerfectPole30}. The correct pole direction is bolded in these tables, and it is observed that these match exactly for the two CoL projections. The bolded pole directions in Tables~\ref{tab:AsteroidPerfectPole60} and \ref{tab:AsteroidPerfectPole30} agree with the simulation truth to machine precision. Also note that the correct (bolded) and incorrect (not bolded) are different enough from each other to allow for easy separation into two groups (i.e., it is easy to match the corresponding pole solutions from the different CoL projections).

\begin{table}[b!]
    \centering
    \caption{Noise-free estimates of pole direction for spacecraft at a latitude of 60 degrees.}
    \begin{tabular}{cccc}
    \hline
     \multicolumn{2}{c}{CoL Projection 1 (red)} & \multicolumn{2}{c}{CoL Projection 2 (cyan)}\\ %\cline{3-4}
     $\hat{\bn}_C$ Sol. 1 & $\hat{\bn}_C$ Sol. 2 & $\hat{\bn}_C$ Sol. 1 & $\hat{\bn}_C$ Sol. 2 \\
    \hline
    \textbf{-0.143087} &  0.098228 & \textbf{-0.143087} &  0.097531 \\
    \textbf{-0.555760} &  0.421967 & \textbf{-0.555760} &  0.419187 \\
    \textbf{-0.818937} & -0.901274 & \textbf{-0.818937} & -0.902646 \\
    \hline
    \end{tabular}
    \label{tab:AsteroidPerfectPole60}
\end{table}

\begin{table}[b!]
    \centering
    \caption{Noise-free estimates of pole direction for spacecraft at a latitude of 30 degrees.}
    \begin{tabular}{cccc}
    \hline
     \multicolumn{2}{c}{CoL Projection 1 (red)} & \multicolumn{2}{c}{CoL Projection 2 (cyan)}\\ %\cline{3-4}
     $\hat{\bn}_C$ Sol. 1 & $\hat{\bn}_C$ Sol. 2 & $\hat{\bn}_C$ Sol. 1 & $\hat{\bn}_C$ Sol. 2 \\
    \hline 
     -0.327999 & \textbf{ 0.392413} & -0.326582 & \textbf{ 0.392413} \\
      0.764579 & \textbf{-0.816434} &  0.761901 & \textbf{-0.816434} \\
     -0.554829 & \textbf{-0.423612} & -0.559331 & \textbf{-0.423612} \\
    \hline
    \end{tabular}
    \label{tab:AsteroidPerfectPole30}
\end{table}

While the correct pole directions in Tables~\ref{tab:AsteroidPerfectPole60} and \ref{tab:AsteroidPerfectPole30} match exactly for each configuration, observe that the incorrect pole directions are slightly different. Unfortunately, however, the incorrect pole directions are usually not different enough for this mismatch to be a reliable means of identifying the incorrect pole in the presence of noise. Additional contextual information (e.g., occlusion check) is necessary to identify which pole solution to use.

\begin{figure}[b!]
    	\centering
    	\includegraphics[width=0.5\textwidth]{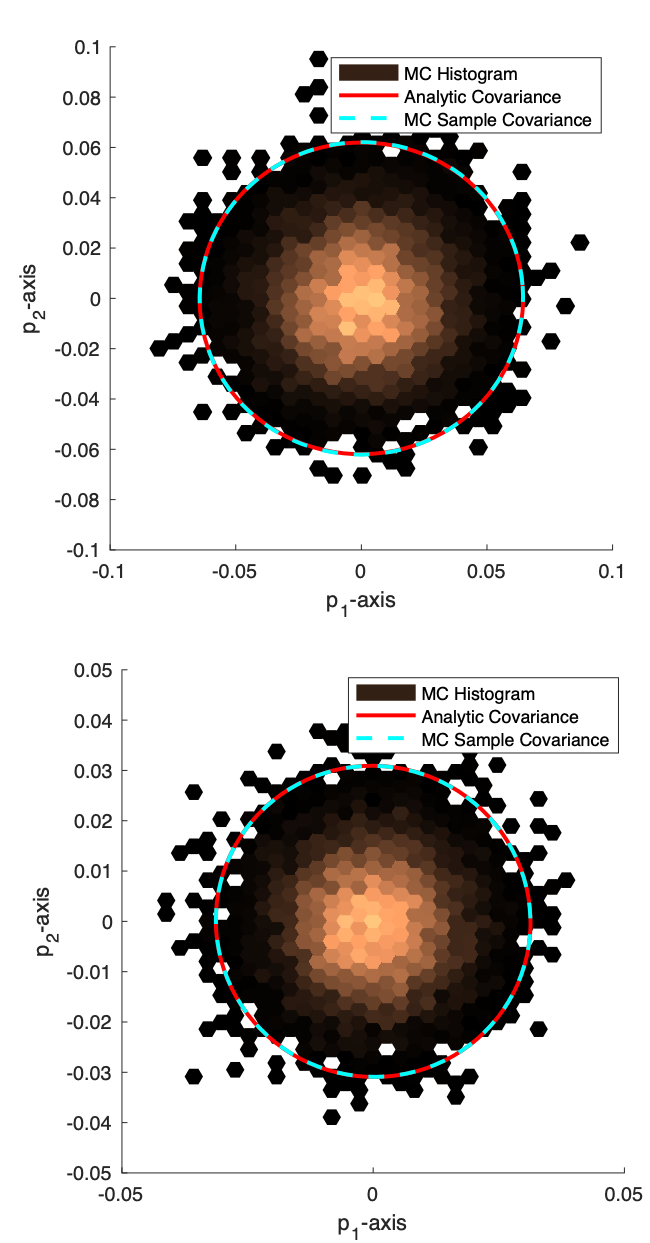}
    	\caption{Pole direction errors in the plane normal to $\hat{\bn}_C$ for a spacecraft at a latitude of 60 degrees (top) and 30 degrees (bottom).}
    	\label{fig:BennuCovPDF}
\end{figure}

In the presence of measurement noise the ellipse fits will not be perfect, and so the estimate of $\hat{\bn}_C$ from Eq.~\eqref{eq:UnitNormalFinal} will not be perfect. In this case, the analytic covariance from Section~\ref{Sec:PoleCovariance} may be used to quantify the errors in $\hat{\bn}_C$ arising from the fit to the red CoL from Figs.~\ref{fig:BennuCoL} and \ref{fig:BennuProjEllipse}. The ellipse fit is performed using the semi-hyper ellipse fit method from Ref.~\cite{Kanatani:2011}. 
Points around image ellipse are sampled with a measurement error of about 15 arcsec. 
Since the semi-hyper ellipse fit directly estimates the implicit coefficient vector $\ba$, the the covariance of the ellipse fit is $\bR_a = E[\delta \ba \, \delta \ba^T]$. The geometric ellipse parameters (e.g., semi-major and semi-minor axis, center coordinates) need not be computed at any point in the process. 
Since the pole estimate covariance $\bR_{\hat{\bn}_C}$ is rank deficient (errors in pole direction lie in the plane perpendicular to $\hat{\bn}_C$), it makes sense to plot the covariance in this plane. This approach is validated for the two configurations shown here via a 10,000-run Monte Carlo analysis and the results are shown in Fig.~\ref{fig:BennuCovPDF}. Excellent agreement is seen between the analytic error covariance (red line) and the Monte Carlo sample covariance (dashed cyan line).

\begin{figure}[b!]
    	\centering
    	\includegraphics[width=0.4\textwidth]{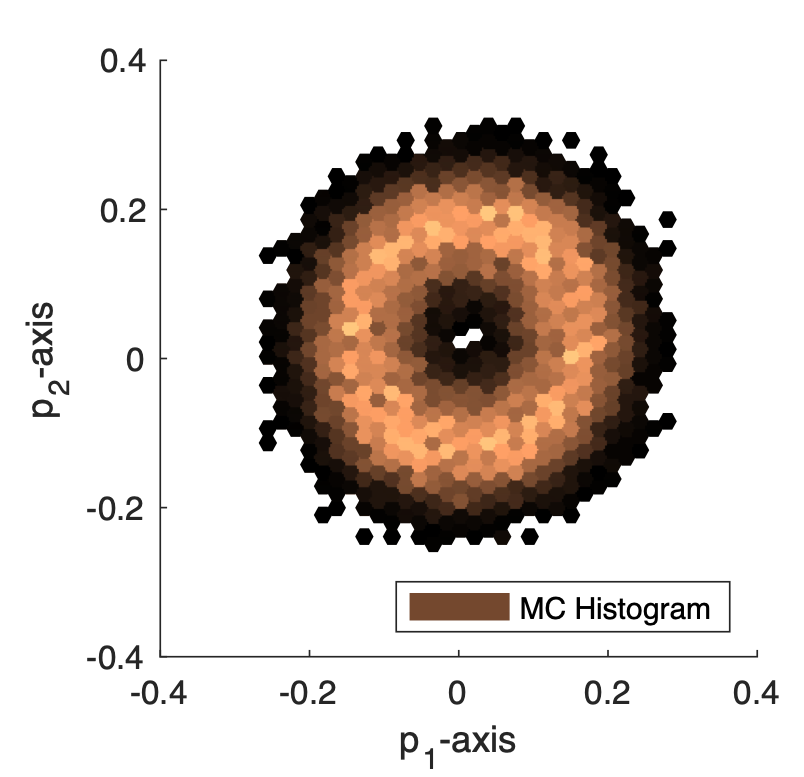}
    	\caption{Pole direction errors in the plane normal to $\hat{\bn}_C$ for a spacecraft at a latitude of 88 degrees.}
    	\label{fig:BennuPolePDF88}
\end{figure}

Pole estimate errors remain approximately Gaussian and the analytic error covariance remains a consistent with Monte Carlo results everywhere except for very high latitudes (i.e., when the camera position is near the line coinciding with the pole). In this case, the projected ellipses become nearly circles and small perturbations in the measured ellipse produce complicated changes in the apparent pole direction. This limitation was discussed in Section~\ref{Sec:PoleCovariance}. To quantify what happens in such a situation, consider a spacecraft at a latitude of 88 degrees. As before, a 10,000-run Monte Carlo analysis is performed. The resulting normal direction errors are as shown in Fig.~\ref{fig:BennuPolePDF88}, which is clearly non-Gaussian. Also note that the magnitude of the errors in the normal direction become larger as the latitude increases, with the worst performance occurring at (or near) a latitude of 90 deg. Consequently, it is desirable to avoid high-latitude observations when using CoL projections to determine the orientation of a spinning body's pole.

\subsection{Example with Jupiter Latitutde Bands}
As the second numerical example, consider the scenario of image-based navigation relative to Jupiter---as previously explored by Danas Rivera and Peck \cite{Rivera:2020,Rivera:2022,Rivera:2023}. As a gas giant, Jupiter's shape is very well-approximated as an oblate spheroid. This simulation uses the 2015 shape model from the International Astronautical Union (IAU) \cite{Archinal:2018}, which defines Jupiter's equatorial radius as $R_E = 71,492 \pm 4$ km and polar radius as $R_P = 66,854 \pm 10$ km. This equates to an ellipticity of $\epsilon = R_P/R_E \approx 0.935$.

Suppose now that a spacecraft collects a single image of Jupiter. In that image, two latitude bands are observed at latitudes of 7 deg and 45 deg. Further suppose that no onboard ''map'' of the latitude bands exists, and so the spacecraft has no internal knowledge of the actual latitude of the observed bands. Instead, the spacecraft only knows that each of these bands correspond to a CoL lying on the surface of an oblate spheroid of known shape (e.g., the 2015 IAU Jupiter shape model). The existing algorithm of Danas Rivera and Peck \cite{Rivera:2020,Rivera:2022,Rivera:2023} could not navigate since it requires correspondence between observed CoL projections and known latitude bands (assumption \#1 in Ref.~\cite{Rivera:2023}). 

To create some representative numerical results, let the spacecraft observation occur from a latitude of 60 deg. Suppose that the camera is pointing in the general direction of Jupiter (so that it appears in the image), but that the camera boresight is not perfectly pointed at Jupiter's center or at the center of either of the CoLs. In such a situation, the algorithm of Danas Rivera and Peck \cite{Rivera:2020,Rivera:2022,Rivera:2023} would not work since the camera is not perfectly pointed at the planet center (assumption \#3 in Ref.~\cite{Rivera:2023}) and the planetocentric distance is not known (assumption \#5 in Ref.~\cite{Rivera:2023}).

\begin{figure}[b!]
    	\centering
    	\includegraphics[width=0.75\textwidth]{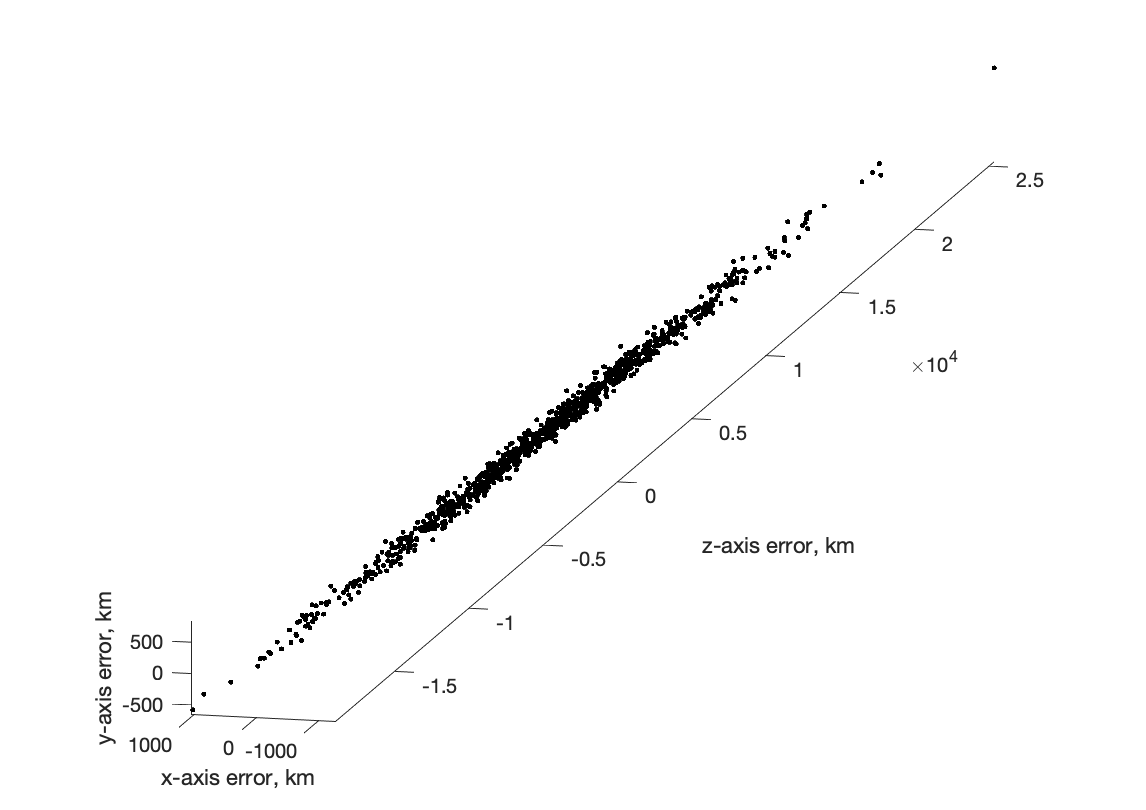}
    	\caption{Scatter plot of Jupiter localization error in the camera frame ($+z$ along boresight).}
    	\label{fig:JupiterError}
\end{figure}

In this case, the orientation of Jupiter's pole within the camera frame may be estimated as described in Section~\ref{Sec:PoleDirection} and similar behavior to Section~\ref{Section:BennuExample} is seen. Now, since Jupiter is an oblate spheroid, the relative position may be found using the results from Section~\ref{Sec:ScaleForSpheroid}. As in the prior example, assume an imaging system producing LOS errors to points around the observed CoL with an error of about 15 arcsec. If the spacecraft views the two specified CoLs from a range of about $50 R_E \approx 3.57 \times 10^6$ km, then Jupiter-relative localization results similar to Fig.~\ref{fig:JupiterError} are obtained. Note that the results in Fig.~\ref{fig:JupiterError} provide the position vector from the spacecraft to Jupiter's center as expressed in the camera frame. As with most vision-based OPNAV techniques, the error is largest in the direction of the observed object. Here, the $1\sigma$ error corresponds to about 1.6\% of the range to Jupiter. 

Since the inertial orientation of navigation cameras is usually well known, this relative position may often be converted to the inertial frame. Thus, assuming the spacecraft inertial orientation is known (e.g., from a star tracker), the CoL projections can be used to find the absolute relative position between the spacecraft and planet---not just the latitude. In this regard, navigation by latitude bands (i.e., by CoLs) produce a very similar measurement as the classical horizon-based OPNAV approach \cite{Christian:2021d}. This flexibility in measurement source (horizon or latitude band) may provide valuable operational flexibility in some situations. In most cases, however, horizon localization will correspond to a crisper feature and have lower image processing errors---and, therefore, horizon-based OPNAV will often provide better navigation results that latitude band-based OPNAV.

\section{Conclusions}
Inspired by recent work on pole estimation \cite{Kuppa:2022a,Kuppa:2022b} and latitude band-based navigation \cite{Rivera:2020,Rivera:2022,Rivera:2023} with images, this work introduces a new framework for navigation with the perspective projection of circles of latitude (CoLs). These advancements begin by transferring established algorithms for image-based localization with circles, including a comparison of different formulations and a discussion of their equivalence. These historical methods are then expanded by introduction of an analytic covariance expression and development of a localization algorithm for oblate spheroids. The result is an algorithmic framework that more accurately captures the projective geometry and allows for more opereational flexiblity (i.e., has fewer constraints) than existing methods.

\section*{Appendix}
The pole covaraince requires computation of a number of measurement partials. These may all be computed analytically. To begin, consider Eq.~\eqref{eq:PartialsN} for $\bbN$,
\begin{equation}
    \bbN = \begin{bmatrix} \partial \hat{\bn}_C / \partial \blam & \partial \hat{\bn}_C / \partial \bu_1 & \partial \hat{\bn}_C / \partial \bu_3 \end{bmatrix}
\end{equation}
which consists of the individual partials
\begin{subequations}
\begin{equation}
   \frac{\partial \hat{\bn}_C}{ \partial \bu_1 } = \sqrt{\frac{\lambda_1 - \lambda_2}{\lambda_1 - \lambda_3 }} \bI_{3\times 3}
\end{equation}
\begin{equation}
   \frac{\partial \hat{\bn}_C}{ \partial \bu_3 } = \pm \sqrt{\frac{\lambda_2 - \lambda_3}{\lambda_1 - \lambda_3 }}\bI_{3\times 3}
\end{equation}
\begin{equation}
    \frac{\partial \hat{\bn}_C}{\partial \blam} = \bg^T \otimes \bm_C
\end{equation}
\end{subequations}
where
\begin{equation}
    \bg = \frac{1}{2(\lambda_1-\lambda_3)^2}
    \begin{bmatrix}
        \lambda_2 - \lambda_3 \\ 
        \lambda_3 - \lambda_1 \\ 
        \lambda_1 - \lambda_2
    \end{bmatrix} 
\end{equation}
\begin{equation}
    \bm_C =  \sqrt{\frac{\lambda_1 - \lambda_3 }{\lambda_1 - \lambda_2}} \; \bu_1 \mp \sqrt{\frac{\lambda_1 - \lambda_3 }{\lambda_2 - \lambda_3}} \; \bu_3
\end{equation}

Then compute the $9 \times 9$ Jacobian of the eigenvalues and relevant eigenvectors with respect to the vectorized matrix $\bA$, as given by the Eq.~\eqref{eq:PartialsM} for $\bM$
\begin{equation}
    \bM = \begin{bmatrix} \partial \blam / \partial \text{vec}(\bA) \\
    \partial \bu_1 / \partial \text{vec}(\bA) \\
    \partial \bu_3 / \partial \text{vec}(\bA) 
     \end{bmatrix}
\end{equation}
which consists of the individual partials (see Refs.~\cite{Magnus:1985} and \cite{Liounis:2019}),
\begin{equation}
    \frac{\partial \blam}{\partial \text{vec}(\bA)}  = \begin{bmatrix} \bu^T_1 \otimes \bu^T_1   \\
    \bu^T_2 \otimes \bu^T_2  \\
    \bu^T_3 \otimes \bu^T_3 
     \end{bmatrix}
\end{equation}
\begin{equation}
    \frac{\partial \bu_i}{\partial \text{vec}(\bA)}  = 
    \bu^T_i \otimes \left[ \left( \bA - \lambda_i \bI_{3 \times 3} - \bu_i \bu_i^T \bA \right)^{-1} \left(\bu_i \bu_i^T - \bA \right) \right]
\end{equation}
Since $\bA$ is symmetric, it only has six unique entries that depend on the $6 \times 1$ vector of coefficients $\ba = [A,B,C,D,F,G]^T$. Given the entries in $\bA$ from Eq.~\eqref{eq:DefA}, it follows that $\text{vec}(\bA) = \bP \ba$, where $\bP$ is
\begin{equation}
\bP = \begin{bmatrix}
     1&   0& 0&   0&   0& 0 \\
     0& 1/2& 0&   0&   0& 0 \\
     0&   0& 0& 1/2&   0& 0 \\
     0& 1/2& 0&   0&   0& 0 \\
     0&   0& 1&   0&   0& 0 \\
     0&   0& 0&   0& 1/2& 0 \\
     0&   0& 0& 1/2&   0& 0 \\
     0&   0& 0&   0& 1/2& 0 \\
     0&   0& 0&   0&   0& 1
    \end{bmatrix}
\end{equation}
Thus, one finds that
\begin{equation}
    \frac{\partial \hat{\bn}_C}{\partial \ba} = \bbN \bM \bP
\end{equation}
with the explicit expressions for $\bbN$, $\bM$, and $\bP$ given above.

Sometimes it is more convenient to compute the partials with respect to the geometric parameters $\bg = [a,b,x_c,y_c,\phi]^T$. The coefficients $\ba = [A,B,C,D,F,G]^T$ may be written directly in terms of $\bg$ according to \cite{Christian:2021crater}
\begin{subequations}
\begin{equation}
A = a^2 \sin^2 \psi + b^2 \cos^2 \psi \label{eq:ConicImplicitA}
\end{equation}
\begin{equation}
B = 2 (b^2 - a^2) \cos \psi \, \sin \psi \label{eq:ConicImplicitB}  \end{equation}
\begin{equation}
C = a^2 \cos^2 \psi + b^2 \sin^2 \psi \label{eq:ConicImplicitC}
\end{equation}
\begin{equation}
D = -2 A x_c - B y_c
\end{equation}
\begin{equation}
F = -B x_c - 2 C y_c
\end{equation}
\begin{equation}
G = A x_c^2 + B x_c y_c + C y_c^2 - a^2 b^2
\end{equation}
\end{subequations}
which allows one to compute the partials $\partial \ba / \partial \bg$. Explicit computation of these partials is tedious, but straightforward.

\bibliography{sample}

\end{document}